# Finite-Size Effects and Dynamical Scaling in Two-Dimensional Josephson Junction Arrays

J. Holzer[1], R. S. Newrock[1], C. J. Lobb[2], T. Aouaroun[1], and S. T. Herbert[3]

1  Department of Physics, University of Cincinnati, Cincinnati, OH  45221-0011

2  Center for Superconductivity Research, Department of Physics, University of Maryland, College Park, MD  20742

3  Department of Physics, Xavier University, Cincinnati, OH 45207-4111

# Abstract

In recent years many groups have used Fisher, Fisher, and Huse (FFH) dynamical scaling to investigate and demonstrate details of the superconducting phase transition. Some attention has been focused on two dimensions where the phase transition is of the Kosterlitz-Thouless-Berezinskii (KTB) type. Pierson *et al.* used FFH dynamical scaling almost exclusively to suggest that the dynamics of the two-dimensional superconducting phase transition may be other than KTB-like. In this work we investigate the ability of scaling behavior by itself to yield useful information on the nature of the transition. We simulate current-voltage (IV) curves for two-dimensional Josephson junction arrays with and without finite-size-induced resistive tails. We find that, for the finite-size effect data, the values of the scaling parameters, specifically the transition temperature and the dynamical scaling exponent $z$, depend critically on the magnitude of the contribution that the resistive tails make to the IV curves. In effect, the values of the scaling parameters depend on the noise floor of the measuring system.

**I. Introduction**

In certain ideal systems, the two-dimensional (2D) superconducting phase transition in zero magnetic field is of the Kosterlitz-Thouless-Berezinskii (KTB) type. For more than two decades, there has been a great deal of work exploring the details of the KTB transition and whether, in fact, one can be truly observed in physically realizable systems. In the past decade, a new class of superconductors has been added to the mix – the high temperature cuprate superconductors. With their layered structures and highly anisotropic coupling strengths, these systems offer the possibility of quasi two-dimensional behavior. While the nature of the superconducting phase transition in high temperature superconductors is as yet an unsettled issue,[1,2,3,4] several authors[5,6] have published work that purports to show the existence of a KTB phase transition as part of a larger three-dimensional transition mechanism. Others[7] do not observe such a transition and believe that the conditions for it do not exist in these materials.

In 1989, Fisher, Fisher, and Huse (FFH)[8,9] offered a general analysis of a superconducting phase transition in D dimensions using a dynamic scaling argument. Their primary focus was on the behavior of superconducting systems in the presence of magnetic fields, but they pointed out that their scaling also applied in zero field to the KTB transition for D = 2 and for a dynamical critical scaling exponent z = 2. In the years since, some groups[10] have used this scaling approach as one measure of proof of the existence or absence of a KTB transition; in effect, if the properly scaled current-voltage (IV) curves collapse (do not collapse) onto universal scaling curves above and below the transition, then a KTB transition is likely (unlikely) to be present. In those cases, the scaling behavior was offered in support of other, more conventional analysis such as the KTB square root temperature dependence in the exponent of the resistive behavior or the existence of a universal jump in the IV exponent.

Recently, Pierson *et al.*[11,12] published a dynamic scaling analysis of IV data taken on ultra-thin (one unit cell thick) high-temperature superconducting films[7] as well as on prototypical 2D low-temperature superconducting systems in which it is believed that a KTB transition exists and has been observed.[5,10,13] Based *primarily* on the results of their scaling they propose that a re-evaluation of the dynamics of the KTB transition may be in order. In particular, they propose that the dynamical critical exponent z may not be 2, as one would expect for diffusive dynamics in systems which follow the 2D XY model, but may be as high as 5 or 6.

In this paper we look to determine the proper role of dynamic scaling in such systems and for a possible source for the very high values of z that Pierson *et al.* observe in their scaling analysis. In particular, we suggest that it is inappropriate to use evidence of scaling behavior in experimental data as the *primary* support for the existence of a 2D phase transition. In the particular case of the KTB transition, scaling behavior should, in fact, be valid only above the transition temperature where a KT correlation length exists and diverges.[14] Below the transition temperature the correlation length is infinite and so we should not observe scaling.[15] Nonetheless, apparent scaling is often found below the transition temperature in real IV data. Medvedyeva, Kim, and Minnhagen (MKM)[16] have suggested that the specifics of the scaling behavior below $T_{KT}$ is determined by the finite size of the sample rather than pointing to evidence of some new dynamics, as Pierson, *et al.* suggest. In their analysis, MKM point out that, although for any finite-size sample the resistance only truly vanishes at zero temperature, for a sample of fixed size L, and for data within a limited temperature region, the resistance may *appear* to vanish at some nonzero temperature, and z may appear to be greater than 2. Thus, one may be lead to believe that a transition to zero resistance may actually occur for values of $z > 2$, when in fact it does not. MKM have designated such an apparent transition a "ghost" transition. Our results are entirely consistent with those of MKM.

We examine this question using the straight-forward approach of simulating the IV characteristics of two-dimensional Josephson-junction arrays, including finite-size effects. As the IV curves are generated using KTB theory, we would expect that scaling would yield parameters consistent with KTB behavior. Instead we find that the details of the scaling, i.e. the values for z and $T_{KT}$, depend in a critical way on the effective voltage sensitivity of the measuring instruments, a purely experimental parameter. We find the mere fact that scaling can be accomplished with values of z other than 2 insufficient evidence for the existence of an alternative phase transition.

We begin in section II with a brief discussion of the nature of the phase transition in a 2D superconductor and the application of dynamic scaling to such systems. In section III we outline the details of our current-voltage simulations and present the results of our scaling in section IV. We conclude with a discussion of the results in section V.

## II. The two-dimensional phase transition in superconductors

For many years it was believed that many types of phase transitions were not possible in two dimensions. For a superconductor, for example, it was believed that as the temperature dropped, the resistivity might become exponentially small but would never be zero, and no true phase transition would actually occur. There were theoretical predictions about the impossibility of general long range order in two-dimensional systems. The earliest was by Peierls[17] who argued that the thermal motion of long wavelength phonons would destroy conventional long range order in a two-dimensional crystal. The absence of certain types of long range order in two dimensions was rigorously shown by Mermin.[18]

The absence of long range order, however, does not necessarily imply the absence of a phase transition. Such a phase transition would be from a disordered high temperature state to an ordered, but not infinite-range, low temperature state. Kosterlitz and Thouless,[19] and Berezinskii[20] showed that this was indeed correct by showing that "quasi long-range order," the algebraic decay of correlations, could occur. Kosterlitz and Thouless called this *topological long range order* and applied it to two-dimensional crystals, *neutral* superfluids, and XY magnets. They did not apply it to two-dimensional superconductors or the isotropic two-dimensional Heisenberg magnet, where, they believed, the proper conditions for observing the transition could not strictly be met.

Beasley *et al.*[21] and Doniach and Huberman[22] demonstrated that Kosterlitz and Thouless's theory could be extended to superconductors under certain special conditions. It is these *special* *conditions* that concern us here. In order to observe a KTB transition – or a KTB-like transition – certain very stringent conditions must be met. If they are not, the *details* of the transition will not be correct and the phase transition will not occur.

What are those conditions as applied to the case of a superconductor?[23] In bulk superconductors the energy to create a vortex is proportional to the length of the vortex and as a result is always much greater than the available thermal energy. However, in thin superconducting films where the perpendicular penetration depth $\lambda_\perp (= \lambda^2/d)$ can be made much greater than the sample size, the energy needed to create a bound *pair* of vortices is $(2\pi n_s \hbar^2/2m) \ln(r/\xi)$, where $n_s$ is the 2D superfluid density, $\xi$ is the superconducting coherence length and r is the separation between the two vortices. This can easily be on the order of $k_B T$. (For an array we have $2\pi E_J \ln(r/a)$ where *a* is the array lattice parameter and $E_J$ is the Josephson coupling energy.) The energy to create a single free vortex on the other hand is $(\pi n_s \hbar^2/2m) \ln(L/\xi)$, infinite in the thermodynamic limit ($L \to \infty$). Thus for temperatures greater than zero, but still sufficiently low, the sample will contain no free vortices, but rather bound pairs of thermally-generated vortices which cannot be driven by an applied electrical current.

The KTB phase transition occurs when these bound pairs of vortices dissociate; this occurs at the Kosterlitz-Thouless-Berezinskii temperature, $T_{KT}$. These now free vortices may be driven by an applied electric current, yielding a flux-flow resistance. Thus, below the vortex unbinding temperature the dissipation is zero in the limit of zero current. Above $T_{KT}$, the resistance is not zero due to the finite density of free vortices and, as is the usual case for flux-flow resistance, the voltage depends linearly on the current, *i.e.*, the system appears ohmic. (Once again, this is strictly correct only in the limit of zero current, as discussed below.) The magnitude of the resistance depends on the density of free vortices, $n_f$, which in turn varies as $1/\xi_+^2$ where $\xi_+$, the correlation length, is a measure of the size of the fluctuations above the transition temperature.

An externally applied current may unbind a pair of vortices via the Lorentz force. Well above the transition temperature, where many pairs of vortices are already unbound, the additional effect of a *small* current unbinding vortex pairs is not observable. That is, above $T_{KT}$ and at low currents the current voltage characteristics are linear due to the thermally unbound vortices. As the current increases and the additional density of free vortices begins to be important, the IV characteristics switch to a power law, $V \propto I^{a(T)}$ where $1 < a(T) < 3$. Below the transition temperature, where there are no thermally unbound free vortices, current unbinding is *always* important, and the current-voltage relations are *always* power law, with $a(T) > 3$. The result is that for *sufficiently small measuring currents,* the exponent of the IV characteristics jumps discontinuously from 1 to 3 at $T_{KT}$.

Whether a KTB transition or KTB-like transition is observable in a particular experimental system depends on the relationships among several length scales: L, the sample size, $\xi_+$, the correlation length for $T \geq T_{KT}$, $\xi_-$, the characteristic size of a bound vortex pair below $T_{KT}$, $r_c$, a critical distance between the two members of the bound pair, and $\lambda_\perp$. The existence of a correlation length is necessary to the scaling we discuss below. $\xi_+$, the size of fluctuations (vortices) above the transition temperature, is a *true* coherence length - true in the sense of point-to-point correlations of the order parameter. $\xi_-$, which may be thought of as the typical separation between the two vortices in a bound pair,[24] is not a true coherence length in that sense. However, the temperature dependence of these two lengths is the same, differing only by a constant.

In general, to experimentally observe a KTB phase transition we must be in the thermodynamic limit – L must be very large. Second, in a superconductor we also require that $\lambda_\perp \gg L$ so that the vortex-vortex interaction is always logarithmic. Finally, we must be in the low current limit to avoid having too many current-unbound vortices above $T_{KT}$. These conditions are most often met in high resistance granular low temperature superconductors and in Josephson-junction arrays – *i.e.*, in weakly-coupled systems.

The problem that most often arises in an experiment is that two of the limits, L very large and $\lambda_\perp \gg L$, are violated – either because the sample is too small or because $\lambda_\perp$ is strongly temperature dependent and crosses over to become smaller than the sample size as the temperature is lowered. In both cases it becomes energetically possible for free vortices to form at all temperatures (either $2\pi E_J \ln (L/a)$ or $2\pi E_J \ln (\lambda_\perp/a)$ is no longer much greater than $k_B T$). These additional free vortices, called finite-size-induced free vortices, are most noticeable well below the transition temperature at very low currents, where they create a linear or ohmic "tail" on the IV characteristics. As we will see, these finite-size-induced free vortices have profound effects on naive dynamic scaling analysis.

*Scaling:*

In a phase transition, sufficiently close to the transition temperature, critical fluctuations are observed. The closer one gets to the transition temperature, the longer these fluctuations will last, and the larger the relevant length scale becomes. In a superconductor the relevant length scale is the coherence length $\xi$. Without loss of generality we can assume that the lifetime of the fluctuations, $\tau$, varies as

$$\tau \propto \xi^z \qquad (1)$$

which defines $z$, the critical exponent. Time "slows-down" as $T \to T_c$. As we approach the critical region, all the physics that really matters is in the diverging length and time scales. In the KTB transition z is expected to be 2. Ammirata *et al.*[11] have suggested that z is considerably larger than 2 in such systems, perhaps as large as 5 or 6. They base this conclusion on a scaling analysis of several experimental systems, some of which[10,13] have heretofore been assumed to display a KTB transition.

FFH presented a generalized scaling law for D-dimensional superconductors using dynamical scaling arguments. We write their result in the following way, using the experimentally determined quantities V and I:

$$V = I\xi^{D-2-z} \rho_\pm \left( \frac{I\xi^{D-1}}{T} \right), \qquad (2)$$

where $\rho_\pm$ are scaling functions above and below the transition whose argument is a dimensionless combination of I, $\xi$, and T. Note that using Eq. (2) implies that correlation lengths exist above and below the transition temperature. In ordinary superconductors this presents no problems. However in thin films and in Josephson-junction arrays this does present a problem since the correlation length is not well defined below the transition; this is further discussed below.

In the rest of this paper we will focus on two-dimensional systems and we rewrite Eq. (2) as

$$V = I\xi^{-z}\rho_{\pm}(I\xi/T). \tag{3}$$

We can remove a factor of $(I\xi/T)^z$ from $\rho_\pm$ and rename it $P_\pm$, yielding

$$\frac{I}{T}\left(\frac{I}{V}\right)^{\frac{1}{z}} = P_{\pm}\left(\frac{I\xi}{T}\right). \tag{4}$$

This is the form often preferred[11,25] for analysis since the coherence length, which tends toward infinity as the transition temperature is approached from above (or below in some systems), only appears in the argument of the scaling function.

At the critical temperature the voltage is proportional to $I^{a(T)}$ for a two-dimensional superconductor, where[26] $a(T) = z + 1$. This results from the coherence length going to infinity as T approaches $T_c$ while the voltage is finite for non-zero currents. However, this power law behavior is also valid for any temperature and current that makes the argument of $\rho_\pm$ tend toward infinity since this drives Eq. (2) to the same limiting form. Thus for high currents V should be a power law function of I at least until other physics enters, *e.g.*, the critical current of the film or junctions is exceeded and the IV curves should once again become ohmic.

Above the transition, in the limit $I\xi/T$ goes to zero, if we assume that the power of the exponent is greater than 1, then we can take $\rho_+ \approx$ constant and

$$\frac{V}{I} = R_{Linear} \propto \xi^{-z}. \tag{5}$$

This is valid for $T > T_c$ and $I \to 0$ and is simply the Kosterlitz-Thouless result just above the transition due to vortex unbinding. Below the transition we cannot take the same limit as it leads to the unphysical result of voltages in the superconducting state.

At this point, all that is required to do dynamic scaling is the temperature dependence of the correlation length. In KTB theory the correlation length $\xi_+$ can be defined above the transition as the size of a fluctuation (i.e. a vortex), or, alternatively, as the average distance between two free vortices. It has a temperature dependence given by,[23]

$$\xi_{+(-)} \propto \exp\left\{\left(\frac{b_{+(-)}}{|T - T_{KT}|}\right)^{1/2}\right\}, \tag{6}$$

where $b_+$ ($b_-$) is a constant of order one. The constant of proportionality depends on the system; for 2D Josephson junction arrays it is essentially *a*, the lattice parameter, while for 2D films it is the Ginsberg-Landau coherence length. Below the transition the correlation length is infinite, and so we often use the quantity $\xi_-$, the typical separation of a *bound* vortex pair, to discuss the dynamics of the vortices. This is not a true correlation length in that it does not come from a two-point correlation function. However, $\xi_-$ is often treated *as if it were* a correlation length, even though it is incorrect to do so, since it has the same temperature dependence as $\xi_+$ to within a constant of order one ($b_+$ and $b_-$ differ by a factor of $2\pi$). To compound the confusion, we often take the temperature dependence of $\xi_+$ and $\xi_-$ to be symmetric for the sake of simplicity. (Indeed, we shall follow this convention for the rest of the paper, i.e. we will assume $b_+ = b_- = b$.) Nevertheless, the fact that the correlation length is not well defined below the transition will have consequences as regards to the scaling behavior and will be further discussed below.

## III. IV Curve Details

In this section we discuss our simulations of the current-voltage characteristics of Josephson-junction arrays. We use standard results from the literature[23] for the power law IV characteristics and for the flux-flow resistance immediately above the transition. Since this system is inherently two-dimensional and theoretically displays a KTB transition in the ideal limit, we should expect any dynamic scaling to yield values consistent with the KTB results,[27] namely $z = 2$. Next we add the voltages caused by finite-size-induced vortex nucleation above and below the transition.[28] We then use dynamic scaling to study these simulations and to determine the effects of finite samples.

Above the transition temperature $T_{TK}$, thermally generated free vortices add a flux flow resistance of the form,[23]

$$R(T)_{t-u} = \frac{V_{t-u}(T)}{I} = 2R_o \frac{L}{W} b_1 \exp\left\{-\left(\frac{b_2}{\tilde{T}-\tilde{T}_{KT}}\right)^{1/2}\right\}, \qquad T > T_{KT} \qquad (7)$$

where t-u stands for "thermally unbound," $\tilde{T} = k_B T/E_J(T)$ is the reduced temperature, $R_o$ is the normal state resistance, L/W is the length/width of the array, and $b_1$ and $b_2$ are constants of order one. (Note that $b_2$ is related to $b_+$ in Eq. (6).) For $T \leq T_{TK}$ this thermally unbound flux flow resistance will be zero since there will be no thermally unbound vortices. In addition to the thermally generated voltage, any finite current will unbind vortex pairs, yielding a voltage of the form,[23]

$$V_{c-u}(T) = 2R_o^{3/2} La \left(\frac{2\pi}{\Phi_o \alpha}\right)^{1/2} [i_c(T)]^{1/2 - \pi E_J(T)/k_B T} [i]^{\pi E_J(T)/k_B T + 1}, \qquad (8)$$

where c-u stands for "current unbound," $a$ is the array lattice parameter, $i_c$ is the critical current per junction, $i$ is the current per junction (roughly I/W), and $\alpha$ is a constant of proportionality. This expression is valid both above and below the transition temperature. Looking at Eq. (8), we see that we can write it as $V_{c-u} \propto I^{a(T)}$, where

$$a(T) = \frac{\pi E_J(T)}{k_B T} + 1, \qquad (9)$$

thus yielding the familiar KTB power law dependence. Above $T_{KT}$, the IV exponent a(T) is given by $1 \leq a(T) < 3$ but the IV curves also have a low-current flux-flow voltage arising from the thermally unbound vortices (see Eq. (7)). Below $T_{KT}$, $a(T) \geq 3$ and the IV curves are pure power law. The total voltage signal is closely approximated as the sum of Eqs. (7) and (8),

$$V = V_{c\text{-}u} + V_{t\text{-}u}. \tag{10}$$

Figure 1a) shows the IV curves generated from Eq. (10), plotted on a log scale to show the power law behavior. Here we used a square array (L = W = 300*a*) with a lattice parameter *a* = 10 µm, a normal state resistance $R_o$ = 100 mΩ, and $T_{KT}$ = 2.55 K. We determined $i_c$ from the universal relation $i_c(T_{KT})$ = (26.706 nA/K)$T_{KT}$ = 68 nA and then calculated $E_J = \hbar\, i_c/2e$. For ease of calculation we suppressed the temperature dependence of $i_c$ (and hence $E_J$), a reasonable approximation near the transition and in the weak Josephson-coupling limit. We also ignore the renormalization correction, which is assumed to be small.

Even though the Eqs. (7) and (8) contain the array size in their expressions, the data of Fig. 1a) assume that we are in the thermodynamic limit (L → ∞). A finite-size sample will contain a population of thermally-generated free vortices both above and below $T_{KT}$. If we assume that $\lambda_\perp$ > L, this finite-size-induced free vortex density can be written as,[28]

$$n_f(T) = \frac{b_3}{a^2} e^{-\pi E_J/k_B T} \left(\frac{L}{a}\right)^{\pi E_J/k_B T}, \tag{11}$$

and the flux flow voltage contributed by these vortices will be of the form

$$V_{f\text{-}s} = a^2 W R_o n_f(T)\, i, \tag{12}$$

where f-s stands for "finite size," and $b_3$ in Eq. (11) is nearly constant for small currents. The total voltage for a finite-size array will be given approximately by the sum of Eqs. (7), (8), and (12),

$$V(T,i) = c_1 V_{c\text{-}u}(T,i) + c_2 V_{t\text{-}u}(T,i) + c_3 V_{f\text{-}s}(T,i), \tag{13}$$

where we have made the temperature and current dependence explicit and added the constants $c_1$, $c_2$, and $c_3$, all roughly of the same order, to allow us to adjust the IV curves so that they appear in a current-voltage window that is roughly experimentally accessible. We emphasize that these constants do not change the essential character of the IV curves but rather change where the deflections will appear in current-voltage space. Figure 1b) shows Eq. (13) plotted on a log scale over an abnormally large voltage scale (the usual range is $10^{-10}$ to 0.1 volts) but over a typical current scale. All other generating parameters were the same for Figs. 1 a) and b), including the values for $c_1$ and $c_2$.

## IV. Current-Voltage Scaling Results

We may now analyze the IV data of Figs. 1 a) and b) using scaling, as expressed in Eq. (4). Our approach is to plot $I^{1+1/z}/[TV^{1/z}]$ as a function of the scaling function variable $x = I\xi/T$ and vary the fitting parameters z, $T_{KT}$, and b to achieve the best collapse onto a scaling curve. In practice, we found many values of $T_{KT}$ which gave an acceptable scaling collapse, but there was always a *highest* value above which no collapse could be achieved. We report below those highest values of $T_{KT}$, and the corresponding values for z and b, that yield the best collapse. This method closely mirrors the procedure followed by Pierson, *et al.*[11] For each of the parameters obtained from a scaling collapse, the uncertainty in the reported transition temperature is ±0.01 K and the uncertainty in the value for z is ±0.03.

The fitting parameters $T_{KT}$ and b are contained in the expression for the KT correlation length $\xi_+$, Eq. (6). It is proper to use $\xi_+$ in the scaling analysis of the data in Fig. 1(a) (no finite-size effects) where the thermodynamic limit is assumed, and then only above the transition, since $\xi_+$ is not well-defined below the transition. For the finite-size effect data (Fig. 1(b)), it is not proper to use $\xi_+$ in the scaling analysis *for all temperatures* because the existence of finite-size-induced free vortices presumes that the correlation length is larger than the sample size,[29] taking us out of the thermodynamic limit. In this case, we should substitute L for the correlation length, at least for those temperatures for which $\xi_+ > L$. Nevertheless, we will proceed by using $\xi_{+(-)}$ for our analysis in order to draw a connection with the work of Pierson, *et al.*

Figure 2 shows the scaling behavior of the data of Fig. 1(a) (no finite-size effects). Here the best scaling collapse occurs for $T_{KT} = 2.55$ K, in agreement with the value used to generate the data, and $z = 2$, in agreement with KT theory. Notice that the data above $T_{KT}$ (lower scaling curve) show excellent scaling behavior in that all of the IV curves collapse onto a single scaling curve with no stray data. This, of course, is not surprising in that the data were generated using the KTB model and evaluated in the regime where the KT correlation length is well defined, so that true scaling behavior is expected. Nevertheless, the lower scaling curve of Fig. 2 sets the standard by which scaling curves using experimental data should be evaluated. Figure 2 inset b) shows an expanded view of the scaling curve above the transition.

The data below the transition (upper scaling curve) do not display as good a scaling collapse as above. Data very near the transition (right side of scaling curve) are slightly askew and do not seem to lie along the same curve, and data at lower temperatures do not collapse completely on top of one another. This curve is, in fact, strongly reminiscent of many scaling curves using experimental data[11] that are considered good evidence of scaling behavior. Most experimental data, however, have power-law dependence over only one or two orders of magnitude (on rare occasions as high as three or four) with a rollover to ohmic behavior at high and low currents. Thus, the data is often culled to include only the power-law portion – typically a very short portion of the IV curve – and, as a consequence, the scaling may appear more favorable than it would otherwise. (As an example, in Fig. 2 inset (a) we plot the data below the transition, but truncated to include only IV data above $10^{-9}$ V.) Conversely, here, where the data is as KTB-like as possible, i.e. *pure* power-law IV curves over 10-15 orders of magnitude with a(T) following the expected temperature dependence, we should expect to see the best scaling collapse possible. That we do not is due to the fact that a correlation length is not well-defined below the transition and so no simple scaling behavior should be expected.

In order to show the effect that finite size and experimental limitations have on scaling behavior, we start with the finite-size-induced free vortex data of Fig. 1b) and introduce a voltage cutoff. This voltage cutoff plays the role of the minimum voltage sensitivity or the experimental voltage noise floor for a measurement system. In Fig. 3 we replot the data of Fig. 1(b) with four voltage cutoffs: one at v = $10^{-7}$ V, $10^{-8}$ V, $10^{-10}$ V, and $10^{-12}$ V. Notice that as the minimum voltage or noise floor is reduced, the effect is to include progressively more of the finite-size induced linear tail in the IV data set. As we shall see, this has a dramatic effect on the parameters of the IV scaling function.

Figures 4 through 7 show the scaling curves obtained for the four voltage cutoffs, v =

$10^{-7}$ V – $10^{-12}$ V, respectively. Here we show the best scaling curve using the highest value of $T_{KT}$ for which a scaling collapse would occur. For the v = $10^{-7}$ V cutoff (Fig. 4), we obtain $T_{KT}$ = 2.29 K and z = 2.23, in contrast to $T_{KT}$ = 2.55 K and z = 2 obtained in Fig. 2. We note that the $10^{-7}$ V cutoff, being the highest of the cutoff voltages, allows for very little of the finite-size-induced linear tail to be included in the IV data set. This is reflected in the scaling parameters being comparatively close to those of IV curves without finite-size induced resistance. As the voltage cutoff is lowered, the value of $T_{KT}$ obtained from the scaling procedure progressively decreases and the value of z increases. For the v = $10^{-12}$ V cutoff the values are $T_{KT}$ = 0.84 K and z = 5.9. By adjusting a parameter that is determined by the experimental measurement system (i.e. the noise floor) we can vary the fitting parameters of the scaling collapse. Contrast this behavior with the non finite-size data where for all voltage cutoffs down to v = $10^{-20}$ V (where we stopped), the same values for z and $T_{KT}$ yield the same scaling collapse. Thus, data obtained on the same sample but measured using different measuring systems can yield completely different scaling fits. This simple fact calls into question the practical viability of exploiting the scaling behavior of IV curves to confirm the details of the phase transition in 2D superconductors.

We also point out that the value of the voltage cutoff is somewhat arbitrary for the data that we generated. That is, for the same selection of voltage cutoffs, we could have altered the results of the scaling fits by changing the values of $c_1$, $c_2$, and $c_3$ in Eq. (13) to allow more or less of the resistive portion to appear above (or below) the cutoff. This is akin to the experimental situation where the noise floor of the measuring system is fixed and the coupling strength of the sample determines how much of the resistive tails of the IV curves will be observable.

Neither is the quality of the scaling collapse an indication of the reliability of the scaling fit. In Figs. 4 - 7 we have plotted the scaling curves as lines rather than data points so as to expose any shortcomings in the scaling collapse. We note that the scaling curves look to be quite good, certainly comparable to most experimental data scaling, despite the wide variation of the scaling parameters. In particular, the data above the transition (lower scaling curves) seem to exhibit an especially good collapse in each case. A closer examination, however, reveals a few problems. In Fig. 4 ($10^{-7}$ V cutoff), for the curves above the transition a slight deviation from KTB behavior at the lowest currents is caused by the addition of finite-size-induced free vortices. This deviation, difficult to discern from the unscaled data (Fig. 3) yet clearly manifest in the scaled, prevents a total scaling collapse of the data (compare with Fig. 2). The deviation is also present in Figs. 5 - 7. As the voltage cutoff is lowered, $T_{KT}$ is reduced; this causes more of the IV curves to end up on the scaling curve above $T_{KT}$. The curve becomes "thickened," making it difficult to distinguish the slight flaws in the collapse. Indeed, if we had used data points of only moderate size, we might not even notice the effect. In addition, the shape of the scaling curve changes, becoming more rounded (once again, compare with Fig. 2). For data below the transition (upper scaling curves) the scaling collapse is not good at all. But as the voltage cutoff is lowered and the apparent $T_{KT}$ reduced, fewer IV curves remain: only those at the lowest temperatures which are now suddenly "near the transition." Consequently, the scaling collapse may appear better than it really is.

## V. Discussion and Conclusion

We have demonstrated that the interpretation of dynamical scaling of IV curves in 2D

systems is subtle. In the thermodynamic limit, while scaling exists and is robust *above* the transition, it does not exist below the transition (Fig. 2). The reason is that the 2D correlation length is well-defined above the transition but infinite at and below the transition. The addition of finite-size effects significantly degrades the scaling. Although a scaling curve can be obtained for finite-size effect data, the scaling parameters are significantly altered, particularly the dynamical critical exponent, z. A change in z would nominally point to a change in the vortex dynamics of the phase transition, indicating other than the diffusive behavior of the KTB picture where $z = 2$. While it is certainly not unreasonable to believe that the vortex dynamics of a finite-sized system may be different from an infinite system (perhaps even entirely different), we are skeptical that the scaling collapse alone is sufficient evidence for this change. The fact that we may obtain a range of values for z simply by truncating the finite-size data at various voltage cutoffs makes the actual value of z for most experimental data highly suspicious, at least in the absence of corroborating support from other analytical methods.

Nevertheless, it is intriguing that Pierson *et al.* find a value of $z \simeq 6$ for a variety of 2D systems, both superconducting and superfluid. Rather than pointing to some universality of physics, however, we suspect that this has more to do with the nature of the data collection and the limitations of instrumentation. In particular, we note that the Johnson noise is the universal noise floor for all measurement systems and that nearly all experimental systems are optimized to approach this limit.

We end with a discussion of the scaling behavior below the phase transition. As mentioned, IV data below $T_{KT}$ should not scale, and a careful examination of the scaling curves (Fig. 2 and Figs. 4-7) shows that this, indeed, is the case (at least in comparison to the quality of the scaling collapse in the lower curve of Fig. 2). The data below the transition, however, certainly does show a *tendency* towards a scaling collapse. MKM have dealt extensively with this question for the 2D-XY model with resistively-shunted-Josephson junction dynamics. They point out that because the low temperature phase is "quasi-critical," with $\xi = \infty$, each temperature is characterized by its *own* scaling function. For small values of the scaling variable x, however, the scaling function may be taken to be temperature-independent. For a finite-sized system, they assume an approximate form for the correlation length, $\xi \propto R^{-\alpha}$, where R is the resistance in the limit of zero current and $\alpha$, in general, depends on both L and T. They then demonstrate that should $\alpha$ be a constant, the resistance would vanish at a temperature for which $z = 1/\alpha$. In the KT picture, MKM note that the resistance actually vanishes only at zero temperature (once again, for finite size), but that it could happen that $\alpha$ is approximately constant over a limited temperature regime causing the resistance to *appear* to vanish at some non-zero temperature. If the IV data happen to fall within this limited temperature regime, we would observe an apparent scaling collapse and the apparent vanishing of the resistance at some specific temperature. That is, we may be tempted to conclude that we have evidence for a phase transition. MKM term this type of transition a "ghost transition."

We may make a connection between this "ghost transition" and our voltage cutoff analysis by noting that the IV exponent a(T) is related to the dynamical critical exponent below the transition, $a(T) = z(T) + 1$. For each of the four voltage cutoffs used, we take the value for z obtained in the scaling process (Figs. 4-7) and identify the corresponding temperature for which the IV exponent is $z + 1$. These IV curves are identified by the arrows above each of the cutoff axes in Fig. 3. Notice that each of these curves may be identified as the one where evidence of the low-current resistive behavior first *disappears* above the corresponding noise floor. That is, all IV curves at temperatures below this one are pure power law-like and the ones above show curvature towards an Ohmic slope. This observation is illusory. If we look below the cutoff voltage, each of the IV curves displays Ohmic behavior at lower currents.

In the MKM picture, this ghost transition temperature will change depending on the finite sample size. For a fixed voltage cutoff (noise floor), decreasing the sample size will cause more finite-size-induced resistance to appear above the cutoff, and the ghost transition will move to lower temperature. This is analogous to our picture in which we keep the size of the sample fixed but allow the voltage cutoff to decrease, thereby uncovering more of the resistive character and affecting the ghost transition. In either case, it is obvious that this does not constitute a "true" phase transition and thus, a search for new vortex dynamics is not required.

Strachan, Lobb, and Newrock[25] have recently offered a methodology for determining whether scaling behavior in experimental IV data is truly indicative of a phase transition or, instead, is an artifact of limited voltage resolution. They use the scaling curve from the best fit to the data to extend the IV curves to voltages below the voltage cutoff. Their argument is that a true scaling curve should predict the shape of the entire IV curve, even that below the noise floor. Using the transition temperature derived from the scaling fit, they then examine IV curves of temperatures equidistant from $T_{KT}$, constructing lines tangent to each curve at the same fixed current. The tangent lines are used to determine the concavity of each IV curve. If $T_{KT}$ represents a true phase transition, the IV curves should exhibit opposite concavity above and below the transition. If the IV curves exhibit the same concavity, the scaling fit is an artifact of the limited voltage resolution of the data.

More recently, Pierson *et al.* has presented a numerical study of a 2D lattice Coulomb gas[30] in which they find that $z = 2$, rather than the $z \approx 6$ obtained in their previous analysis. They attribute the inflated value of z to finite-size effects, a result consistent with the results we show here. They also claim to show evidence of the existence of a 2D correlation length below the transition, a result at odds with our analysis.

## VI. Acknowledgments

We would like to acknowledge useful discussions with Petter Minnhagen. This work was supported in part by grants from the NSF under contract No. DMR– 9801825 (Cincinnati) and contract No. DMR–973280 (Maryland), and from the John Hauck Foundation (Xavier).

# Figure Captions

Figure 1: a) Simulated current-voltage curves for a Josephson junction array in the thermodynamic limit (no finite-size effects) for temperatures varying between 0.5 K and 2.76 K, with temperature steps of 0.05 K. The dark line indicates $T_{KT}$. b) Current-voltage curves including finite-size-induced vortices. Temperatures are shown every 0.1 K

Figure 2: Scaled IV curves for data of Fig. 1a) including no finite-size effects. Inset a) shows a blowup of the data above the transition, but with the IV data truncated below $10^{-9}$ V. Inset b) shows a blowup of the scaling curve below the transition to show the details of the scaling collapse.

Figure 3: Replot of Fig. 1b) showing the voltage cutoffs. IV curves are shown every 0.05 K. The arrows denote the location of the "ghost transition" for each cutoff value (see text), the point at which the resistive character disappears.

Figure 4: Scaling collapse of the finite-size-induced resistive IV data (Fig. 3) with a voltage cutoff of $10^{-7}$ V.

Figure 5: Scaling collapse of the finite-size-induced resistive IV data (Fig. 3) with a voltage cutoff of $10^{-8}$ V.

Figure 6: Scaling collapse of the finite-size-induced resistive IV data (Fig. 3) with a voltage cutoff of $10^{-10}$ V.

Figure 7: Scaling collapse of the finite-size-induced resistive IV data (Fig. 3) with a voltage cutoff of $10^{-12}$ V.

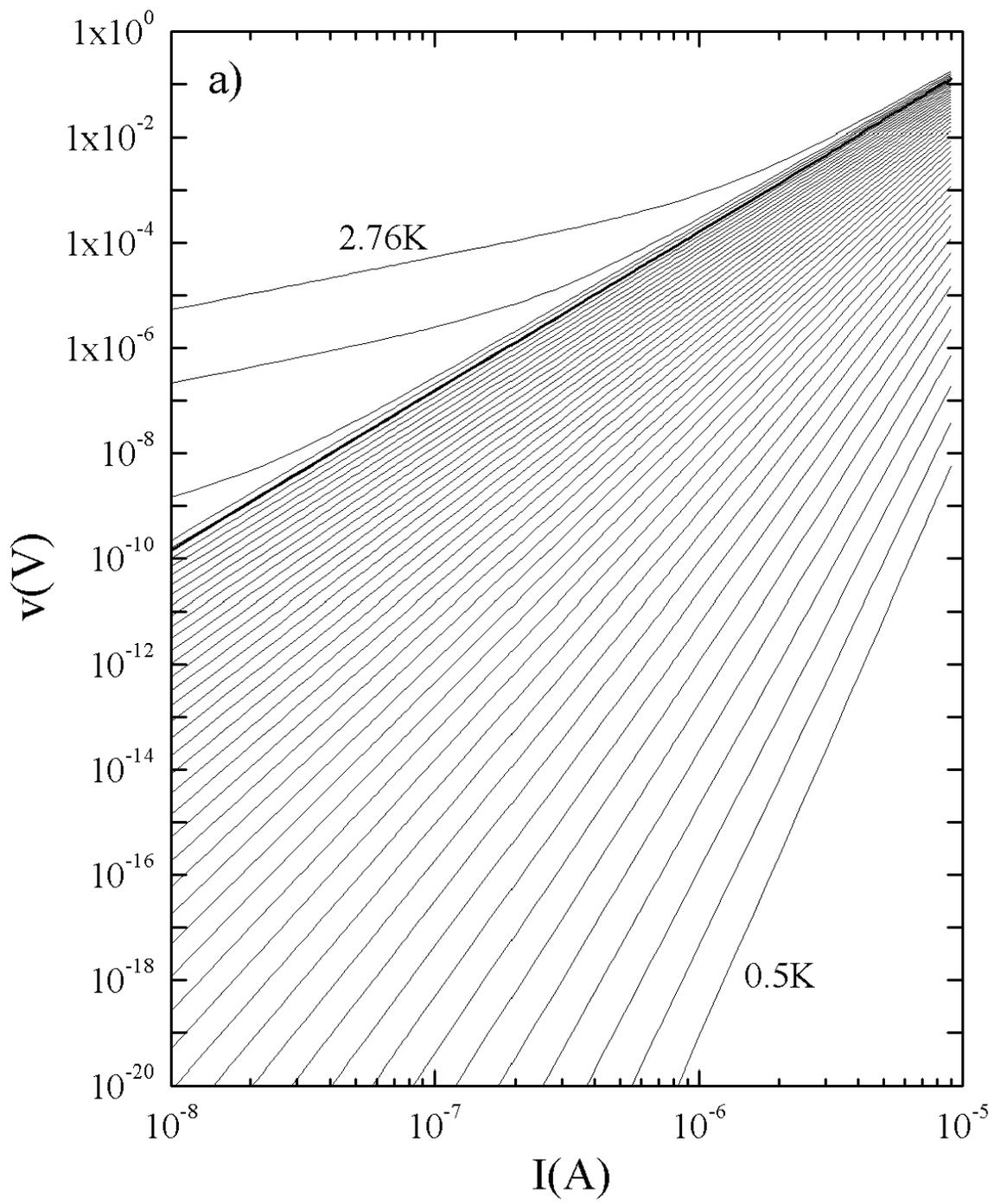

Fig. 1 a)

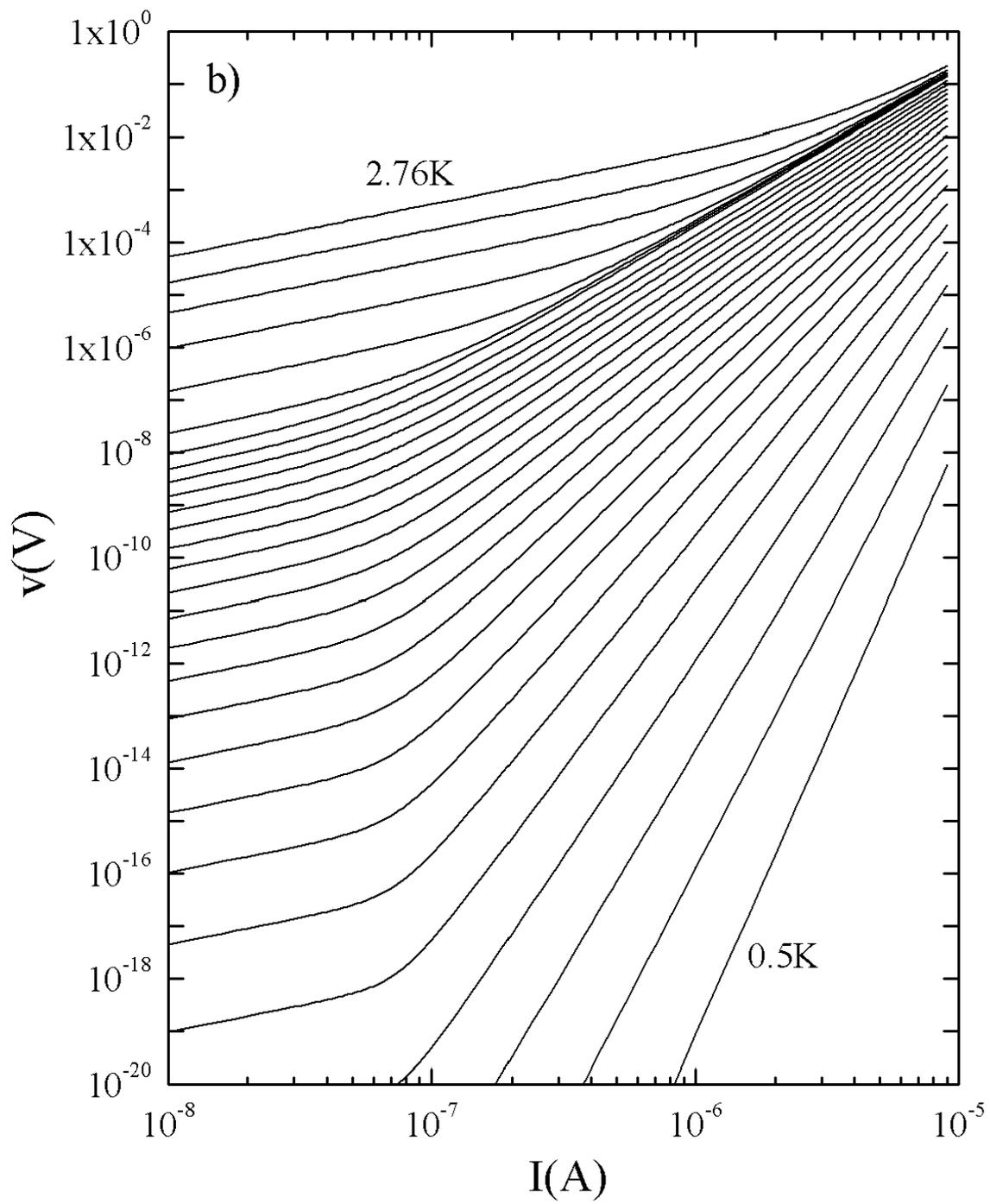

Fig. 1 b)

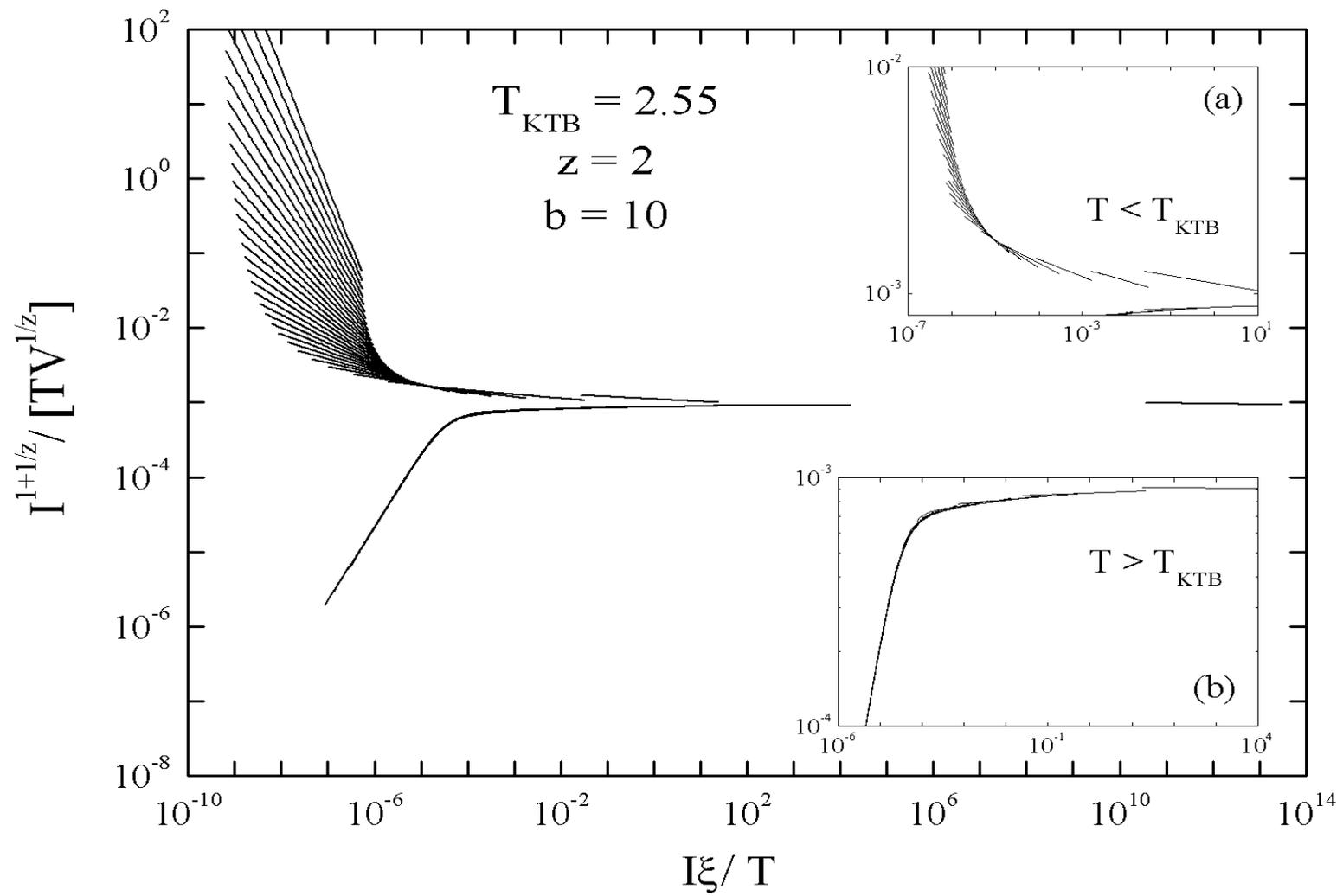

Fig. 2

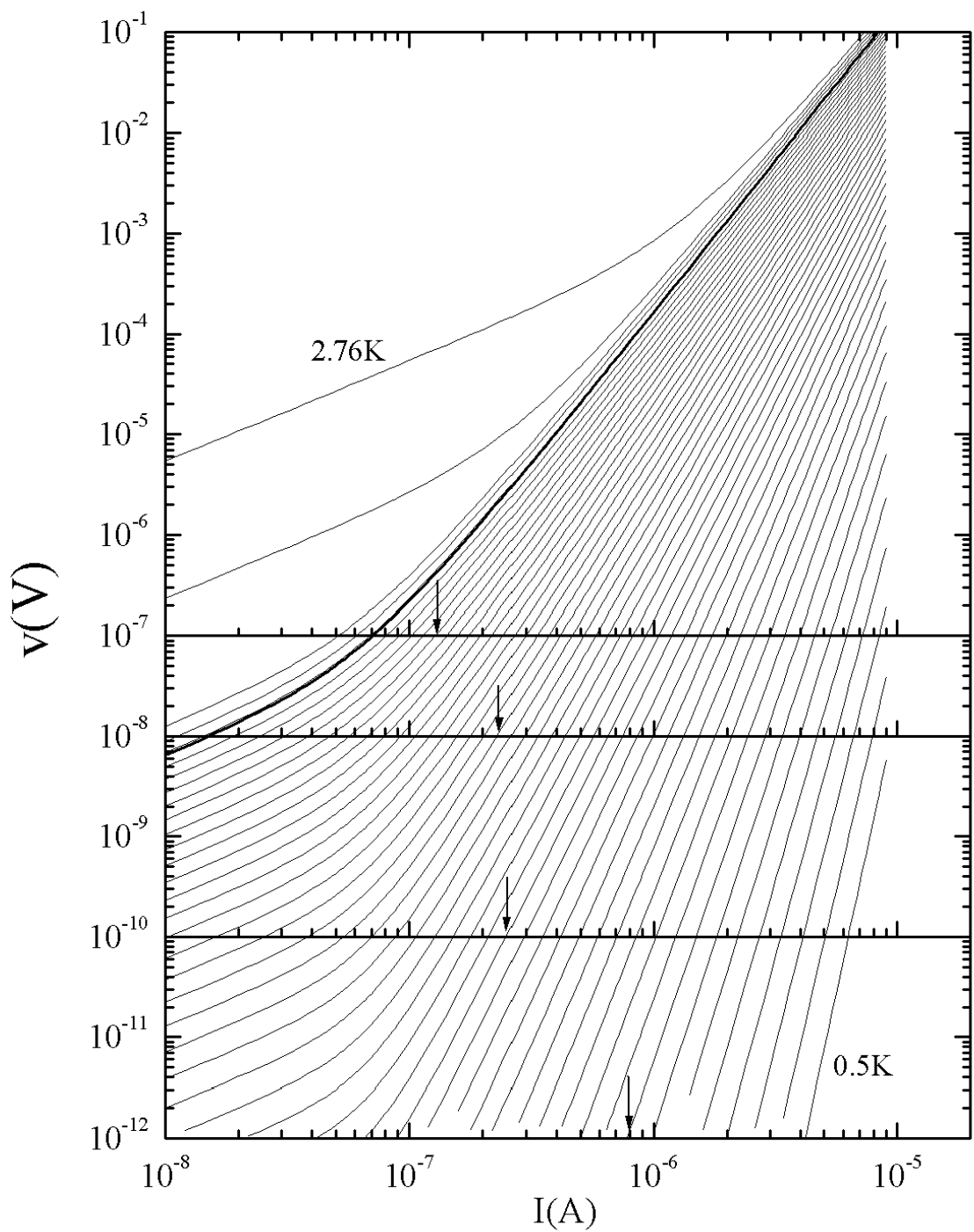

Fig. 3

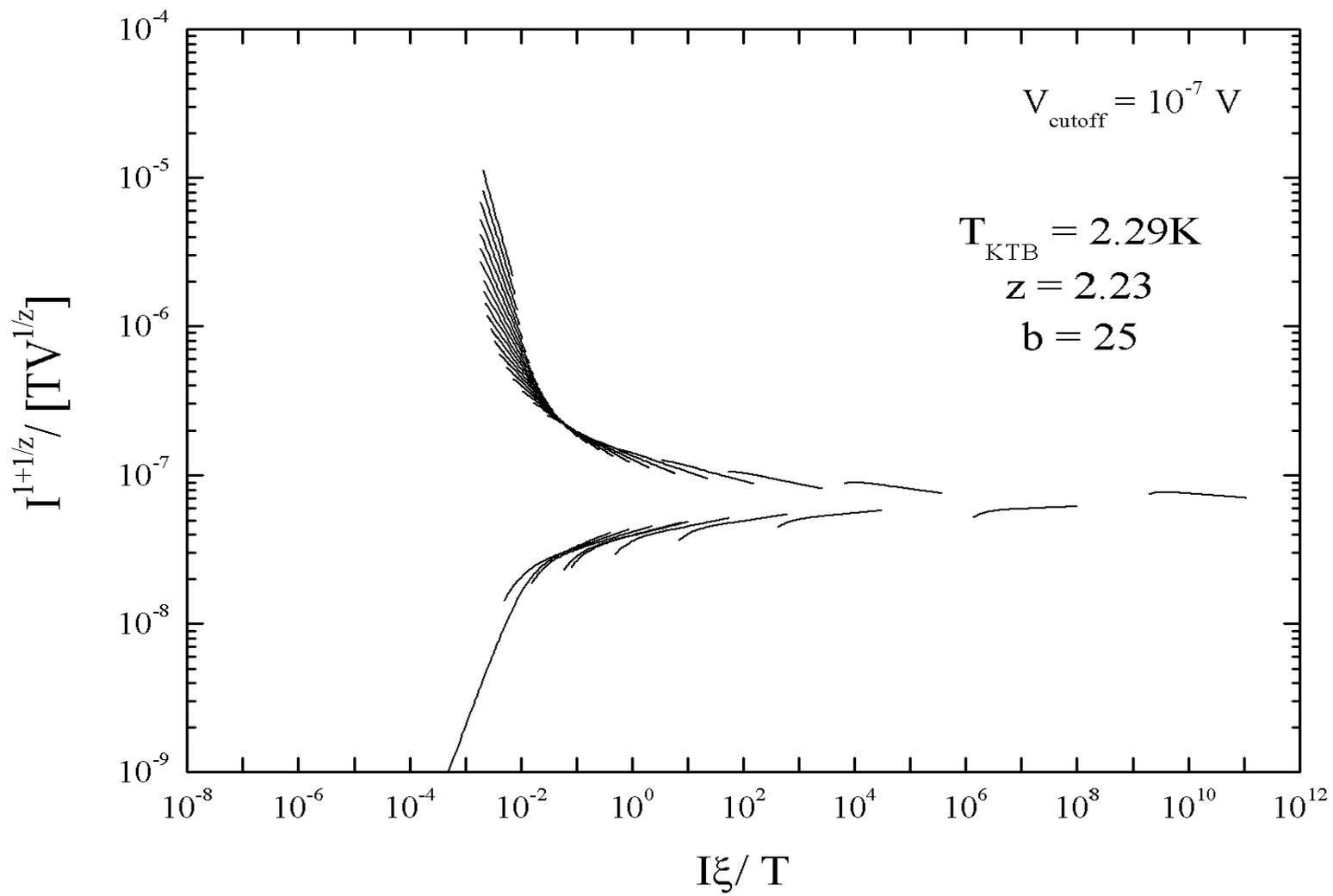

Fig. 4

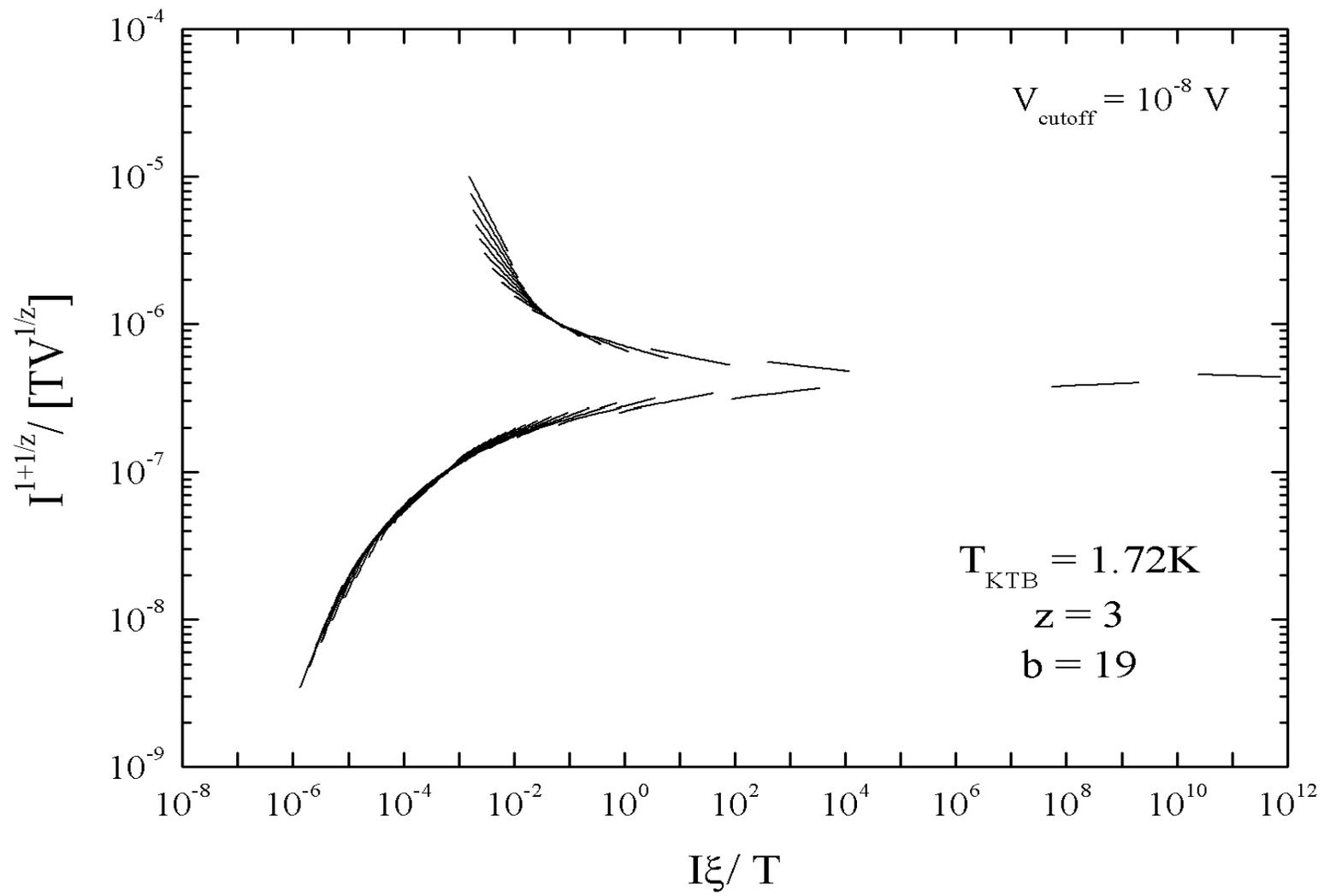

Fig. 5

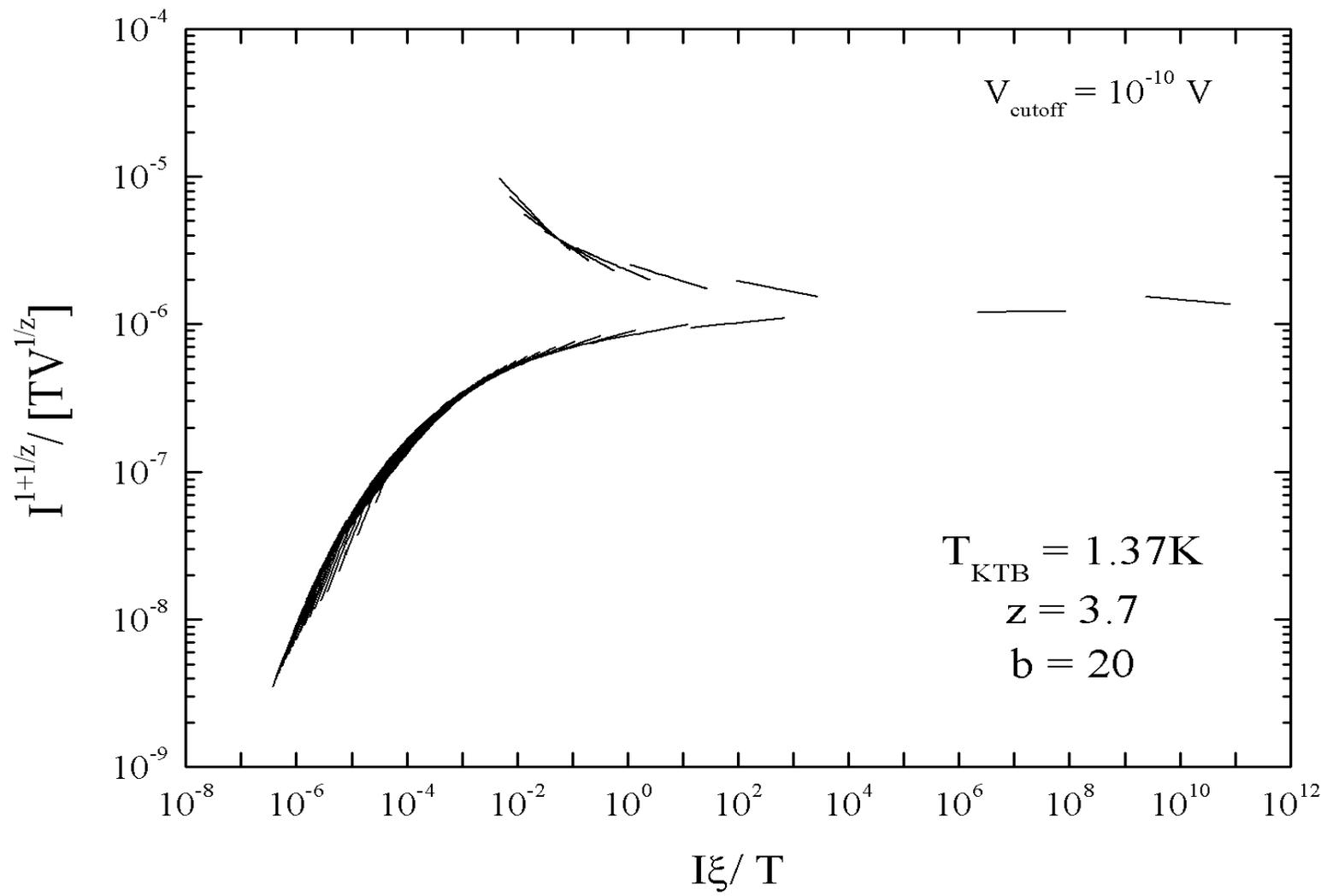

Fig. 6

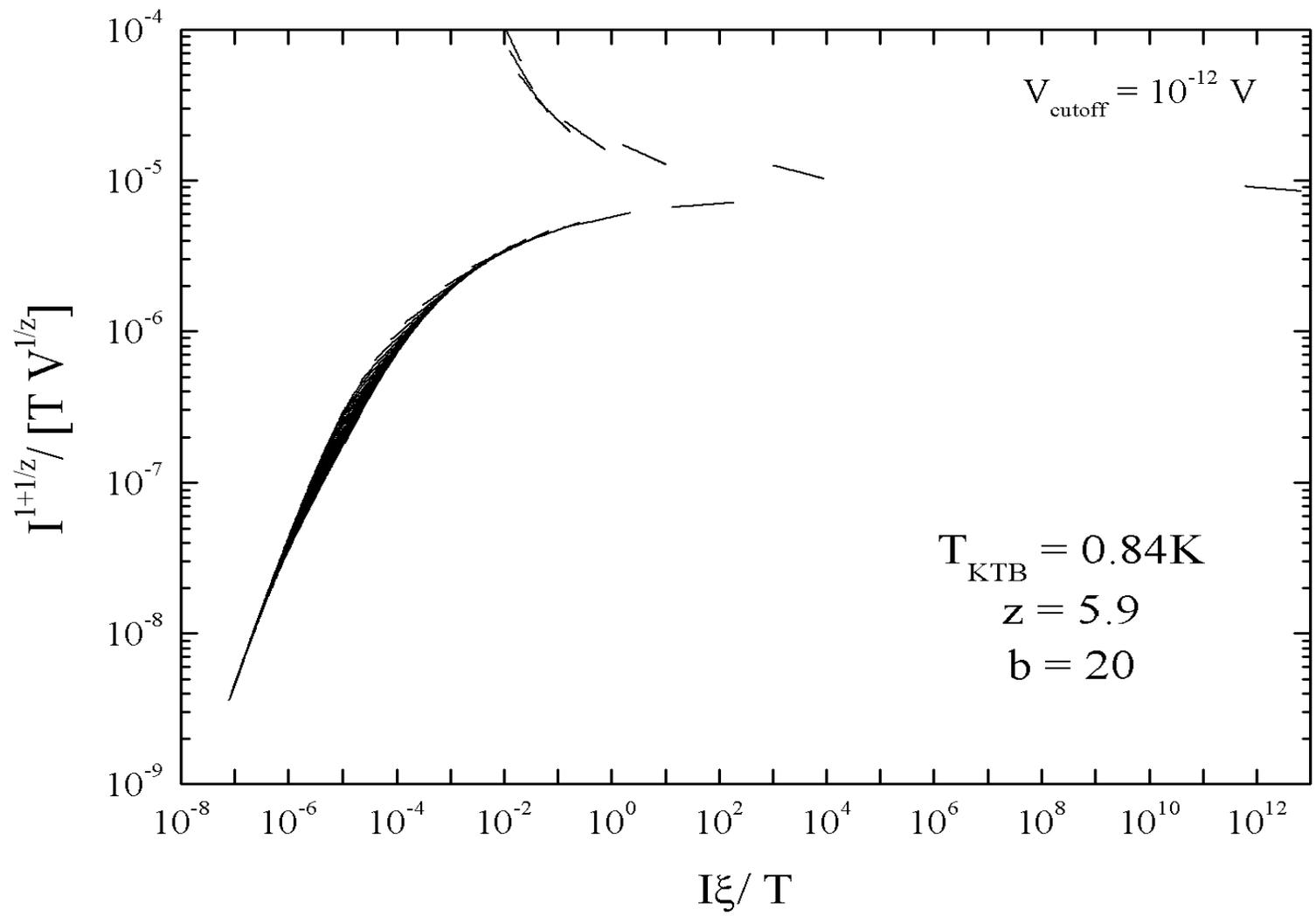

Fig. 7